# Substrate-Independent Catalyst-Free Synthesis of High-Purity $Bi_2Se_3$ Nanostructures

*Jerome T. Mlack[1], Atikur Rahman[1], Gary L. Johns[1], Kenneth J. T. Livi[2], Nina Marković[1*]*

**[1]Department of Physics and Astronomy, Johns Hopkins University, Baltimore, Maryland 21218**

**[2]The High-Resolution Analytical Electron Microbeam Facility of the Integrated Imaging Center, Department of Earth and Planetary Sciences, Johns Hopkins University, Baltimore, Maryland 21218**

ABSTRACT. We describe a catalyst-free vapor-solid synthesis of bismuth selenide ($Bi_2Se_3$) nanostructures at ambient pressure with $H_2$ as carrier gas. The nanostructures were synthesized on glass, silicon and mica substrates and the method yields a variety of nanostructures: nanowires, nanoribbons, nanoplatelets and nanoflakes. The materials analysis shows high chemical purity in all cases, without sacrificing the crystalline structure of $Bi_2Se_3$. Low-temperature measurements of the nanostructures indicate contributions from the surface states with a tunable carrier density. Samples synthesized on flexible mica substrates show no significant change in resistance upon bending, indicating robustness of as-grown $Bi_2Se_3$ nanostructures and their suitability for device applications.

Topological insulators (TIs) are unconventional bulk insulators with non-trivial metallic surface states that arise due to strong spin-orbit coupling[1,2]. These surface states exhibit a Dirac-like dispersion and are topologically protected from scattering on non-magnetic impurities and crystal defects[3-6]. Their unique properties offer exciting possibilities for use in spintronics[7] and quantum information[8] applications, as well as for observing exotic phenomena such as the exciton condensate[9] and Majorana fermions[10]. One of the most intensely studied topological insulators is bismuth selenide ($Bi_2Se_3$), whose rather large band gap (0.3 eV) allows the possibility to access the topologically protected surface states at room temperature. $Bi_2Se_3$ has been synthesized in various forms by molecular beam[12,13] and hot wall epitaxy[14], single crystal growth[15-17], sonoelectrochemical methods[18,19], mechanical exfoliation[20,21], and chemical vapor deposition[22-24]. Angle-resolved photoemission spectroscopy[15,16,25] and scanning tunneling microscopy[22] measurements have shown evidence of the surface states in bulk, thin film, and nanostructures of $Bi_2Se_3$. Nanostructures of topological insulators are especially useful for the transport measurements[20-24,26-33], as their high surface-to-volume ratio tends to minimize the contribution from the bulk impurities and emphasize the conduction through the surface states. At this point, both fundamental research and the practical application possibilities face some challenges involving the sample fabrication. One challenge is to find a robust and simple growth method that guarantees reproducible nanostructures of high purity on a variety of substrates. Another challenge is to separate the conduction through the surface states from the contribution of bulk carriers with energies inside the gap, which typically exist due to inadvertent doping or structural disorder. The bulk carriers also make it difficult to control the carrier density through gating, which is desirable for most applications. In order to increase the purity and minimize the number of bulk carriers due to contamination, synthesis efforts and transport measurements have recently turned to catalyst-free nanostructures[18,19,23,28]. The reported methods require growth under vacuum[23,28], the growth time can be several hours[18,19], and some methods yield only nanoplatelets[23,28]. Here we demonstrate the synthesis of $Bi_2Se_3$ nanostructures using a catalyst-free vapor-solid (VS) method with $H_2$ as carrier gas. All the materials are commercially available and the growth is achieved in a short time period at ambient pressure in a standard tube furnace. Nanowires, nanoribbons, nanoplatelets and nanoflakes of different sizes and shapes are found in abundance on glass, silicon and flexible mica substrates. This method has proved to be easy, fast and highly reproducible, consistently yielding high-quality, ultra-pure nanostructures on a variety of materials such as semiconducting (silicon), transparent (glass), and flexible (mica) substrates, which are useful for various applications. Although bulk carriers are still present, low-temperature transport measurements indicate a significant contribution from the surface states without additional chemical doping or processing.

The $Bi_2Se_3$ nanostructures were grown using the VS method in a quartz tube furnace at ambient pressure. Alfa Aesar vapor deposition grade $Bi_2Se_3$ crystals of 99.999% purity were used as the source material. Source crystals were weighed and placed in a ceramic combustion boat at the center of a quartz tube in a tube furnace, as illustrated in Figure 1(a). Glass slide substrates (Fisher Scientific) were cleaned before growth by sonication in acetone for five minutes, followed by isopropyl alcohol (IPA) for five minutes. The glass substrates were dried using $N_2$, and then placed on a hot plate at 210°C for 90 seconds. The same cleaning procedure was used for silicon substrates. To ensure successful growth on mica substrates, scotch tape was used to peel away the top layers and expose a clean, flat growth surface. The substrates were then inserted into the quartz tube, starting at 13.5 cm from the center, and ranging to 19 cm from the center. Silicon and mica substrates, when used, were placed on top of a glass slide before insertion into the quartz tube, in order to maintain the appropriate height and to keep the flexible substrates flat. Argon gas was allowed to flow through the tube at a rate of 1550 sccm for fifteen minutes in order to purge the tube of oxygen. After purging, the rate was reduced to 200 sccm and $H_2$ gas was added to the Ar flow at a rate of 5 sccm. The temperature was set to 700°C, and after reaching the set temperature, the $H_2$ gas was turned off and the furnace was allowed to cool. The overall growth time in each case was less than five minutes. When removed from the furnace, all substrates had an observable grey coating.

Nanoflakes and ribbons were found to grow on the entire slide from 13.5-16 cm from the center, while nanowires tended to grow in the range of 14-15.5 cm from the center. These areas correspond to temperatures ranging from 530-337°C. Figures 1(b) and 1(c) show optical and scanning electron microscope images, respectively, of the growth results on a glass slide. The images show an abundance of nanostructures including nanoribbons, nanowires, and nanoplatelets. Some nanoplatelets appear to have grown on top of one another, forming taller structures that can be seen on the images. An SEM image of the growth results on a silicon substrate is shown in Figure 1(d) (see Supporting Information for additional images). It is evident that the variety and the size of the nanostructures grown on silicon are very similar to those grown on glass slides. In contrast, the nanostructures grown on mica were largely composite-nanoplatelet films, covering large portions of the substrate, although some regions also contained nanowires and ribbons. An optical microscope image of nanostructures grown on mica is shown in Figure 1(e). The type of the nanostructures was found to depend on the temperature profile of the furnace and could be controlled by positioning the substrate in the furnace (see Supporting Information for more details).

Nanostructure growth was also achieved without $H_2$ as a carrier gas, but the resulting nanostructures are significantly smaller and less abundant (see Supporting Information for more details and images). The most likely reason is that a smaller amount of material is carried from the source to the substrate in the absence of the $H_2$ gas. The $H_2$ gas is known to bind with Se to form $H_2Se$, possibly increasing the amount of material transported downstream to the substrates.

The basic crystal structure and the chemical make-up of the nanostructures grown on various substrates were characterized using a transmission electron microscope (TEM). Figure 2(a) shows a TEM image of a typical nanowire, grown on a glass slide, of approximately 90 nm in diameter. The nanowires have a uniform width and a large aspect ratio, and are typically tens of micrometers long, as shown in Figure 2(b). The crystal structure, as shown in Figure 2(c), was characterized using the high resolution TEM (HRTEM). The image shows an ordered structure with a lattice spacing of 0.21 nm which is consistent with other lattice spacing measurements of $Bi_2Se_3$ nanowires[22]. An image of a typical hexagonal nanoplatelet is shown in Figure 2(d). The selected area electron diffraction (SAED) pattern of the platelet in Figure 2(e) shows normal incidence indicative of a hexagonal structure, with growth occurring along the [11$\bar{2}$0] direction and [001] planes being on the top and bottom. Typical energy dispersive x-ray spectroscopy (EDS) is shown in Figure 2(f). Pronounced Bi and Se peaks are observed, indicating that the composition remains unaltered. Presence of copper is due to the copper TEM grid on which the structures were suspended. The nanostructures do contain some crystal defects and grain boundaries, which are also observed in nanostructures obtained using other methods. However, as our method requires no catalyst, the chemical purity is rather high.

In order to investigate the electrical properties of the $Bi_2Se_3$ nanostructures, nanodevices were fabricated using a combination of optical, electron beam (e-beam), and Focused Ion Beam (FIB) lithography. The $Bi_2Se_3$ nanostructures were removed from the glass or silicon substrates via sonication in IPA. The solution of nanostructures in IPA was then drop-cast and dried on a silicon substrate, coated with 300 nm of $SiO_2$. Standard optical lithography was used to fabricate the contact pads (5nm Cr/55nm Au). In order to ensure good electrical contact to $Bi_2Se_3$, we used the FIB to etch the oxide layer from the surface of $Bi_2Se_3$ and to deposit platinum leads directly on the etched areas, without breaking vacuum (only the areas directly under the contacts were etched). The platinum leads were then connected to the bonding pads using palladium leads, fabricated by standard e-beam lithography and sputtering.

Electron transport measurements were carried out using the four-probe geometry. The current through the sample was reversed to eliminate any thermovoltages and was kept low to avoid heating. Hall resistance was measured on devices with the Hall bar geometry, such as the one shown in Figure 3(a). Low-temperature measurements were carried out using a $^3$He cryostat with a base temperature of 250 mK, equipped with an 8 Tesla magnet. Electrical measurements were done using an analog lock-in amplifier (PAR 124A). Low frequency (17 Hz) bias current was driven through the sample, and the voltage was measured after amplification by a SR 560 low-noise voltage preamplifier (gain set at 100).

The resistance as a function of temperature is shown in Figure 3(b). The resistance was measured by driving 100nA bias current through the current leads, marked 1 and 2 in Figure 3(a). The longitudinal voltage drop was measured on leads 5 and 6 in the four-probe geometry. The sample resistance decreases with decreasing temperature above 14K, below which it increases with decreasing temperature. The initial metallic behavior arises due to excess free carriers generated from the Se vacancies, which moves the chemical potential into the conduction band, and is typically observed in most as-grown crystals of $Bi_2Se_3$. Increasing resistance with decreasing temperature below 14K is an indication of the onset of the insulator-like behavior, expected to arise due to the freezing-out of excess charge carriers. With decreasing concentrations of free carriers, the resistance upturn is expected to take place at higher temperatures. This has been demonstrated by doping $Bi_2Se_3$ with Ca.[38] The observed resistance upturn in our as-grown sample indicates a relatively low amount of free carriers. However, the small increase in the resistance indicates that the contribution from the bulk is still significant. The resistance increases as a function of a magnetic field, as shown in Figure 3(c), which appears to be mostly due to bulk carriers. A measurement of the Hall resistance $R_T$ (current bias on leads 1 and 2, and voltage measured on leads 3 and 5) as a function of magnetic field is shown in the inset of Figure 3(d). A careful inspection indicates that the Hall resistance is not linear (see Supporting Information for more details). This indicates presence of multiple types of carriers with comparable mobility. In Figure 3(d), the Hall conductance is fit using the two-carrier model[34] which gives the carrier concentrations of $n_1$ = 6.0e+13 1/cm$^2$ and $n_2$ = 7.8e+12 1/cm$^2$. These values are of the same order of magnitude as those reported by groups using high-quality MBE growth methods[34]. Both the longitudinal resistance and the Hall coefficient depend on the gate voltage. As shown in Figure 3(e), the Hall coefficient increases and the resistance decreases for both polarities of gate voltage. The change in the Hall coefficient is most likely due to the electrostatic accumulation of charge carriers near the bottom surface (holes for negative and electrons for positive gate voltages) [31,32]. This implies that the Fermi level is near the Dirac point of the bottom surface around zero gate voltage. We were not able to observe the crossover between p and n-type carriers with the gate voltage available in our experiment, but the resistance decreases for both polarities of the gate voltage, as shown in Figure 3 (f), which might be expected if the Fermi level were indeed close to the Dirac point [31,32]. However, the relatively small change of both Hall and longitudinal resistance again indicates the contributions from bulk channels.

Figure 4(a) shows an SEM image of another device, a nanowire that was measured down to 250 mK. Leads 1 and 2 were used as current leads, and the voltage was measured between leads 3 and 4. At low temperatures, the resistance increases with decreasing temperatures, and appears to be saturating at the lowest temperatures, as shown in Figure 4 (b). The resistance also depends on the back gate voltage, and the Dirac point is located at -4 V, as shown in Figure 4(c). Figure 4(d) shows the magnetic field dependence of the conductance, which indicates negative magnetoconductance. Sharp negative magnetoconductance can be an indication of weak antilocalization (WAL), often observed in $Bi_2Se_3$ nanostructures at low temperatures[34]. Even though a reasonable fit to the theoretical expression for WAL corrections can be obtained (see Supplementary Information for more details), fitting the magnetoconductance data in this case is complicated by the sample size and geometry - the dimensionality, universal conductance fluctuations and scattering between the surface and the bulk channels all play important roles [24,35,37]. A more detailed study is needed in the future.

We have also grown $Bi_2Se_3$ nanostructures on flexible mica substrates. Figure 5 shows the results of the $H_2$–assisted catalyst-free growth method on a flexible mica substrate in the 14-16 cm temperature region. Figures 5(a) and 5(b) show different examples of the typical layering of merged triangular and hexagonal nanoplatelets. Such layered nanostructures occur in addition to nanowires shown in Figure 1(e). Figure 5(c) shows an AFM (Atomic Force Microscope) image of a nanoplatelet layer. The vertical thickness of the layering, indicated by the two red arrows, is approximately 3 nm, which corresponds to the **c**-axis height of the $Bi_2Se_3$ unit cell (three quintuple layers)[36].

To determine the mechanical robustness of the $Bi_2Se_3$ structures, a large composite-nanoplatelet film was grown on the flexible mica substrate. Indium electrodes were attached to the nanoplatelet film using silver paint, and the transport properties were measured at room temperature. The substrate was then bent using a vice, as shown in Figure 5(d). Similar film resistance was measured for both strained and unstrained positions as shown in Figure 5(e). Even after several cycles of bending and stretching, the average resistance of the film was found to be 1.5 kΩ for both flat and bent mica substrate. This indicates uniform and homogeneous coverage of nanoplatelets film over the entire substrate.

In conclusion, we have grown $Bi_2Se_3$ nanostructures using a catalyst-free VS method with $H_2$ as a carrier gas. The growth was performed at ambient pressure for a short time period and yielded nanowires, nanoribbons, nanoplatelets and nanoflakes, with a variety of geometries and sizes. The $H_2$ gas plays an important role in achieving the abundance and the variety of $Bi_2Se_3$ nanostructures. Transport measurements on devices made from the nanostructures indicate the presence of two carrier channels with a contribution from bulk. The observed resistance upturn at low temperature, the positive magnetoresistance and tunability of the carrier concentration hold promise for the future studies of surface states. By growing on various substrates, we have shown that this growth method allows for robust device applications.

FIGURES

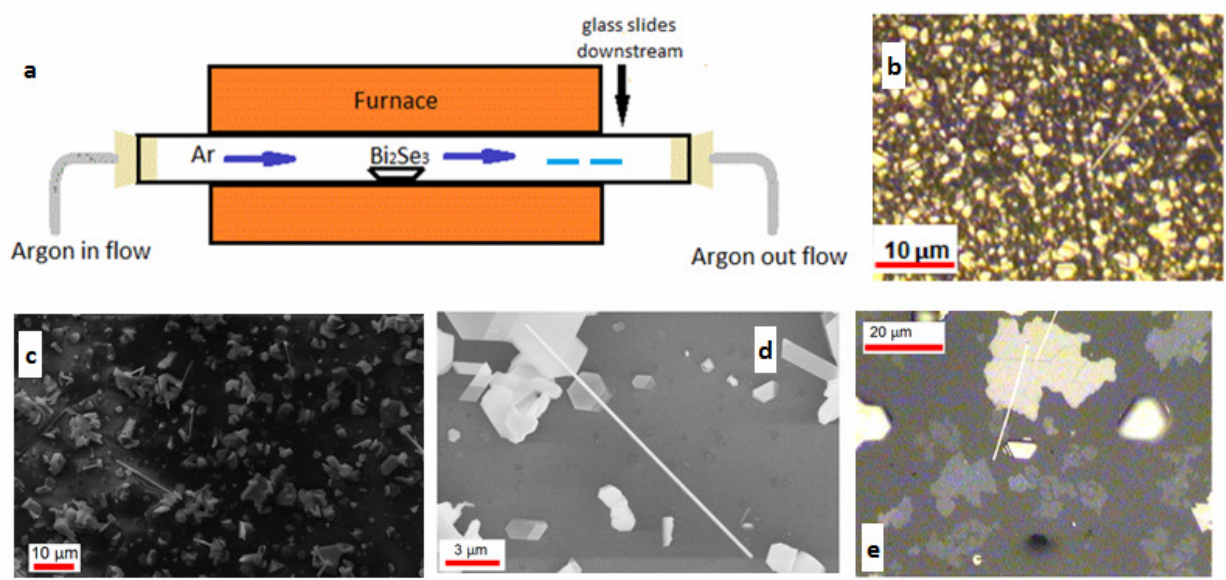

**Figure 1. Diagram of the growth setup and results. (a) Quartz tube furnace setup. A ceramic boat containing $Bi_2Se_3$ is located in the center of the tube. On the right end of the tube are glass slides of size 2.3x2.54 cm. The slides are located in a range from 13.5-19 cm from the center of the furnace. Prior to growth, argon is allowed to flow at a rate of ~1500 sccm for 15 minutes to purge the system of any oxygen. After purging, the flow rate is decreased to 200 sccm, $H_2$ gas flow of 5 sccm is started, and the furnace is set to 700°C. When the furnace reaches the set temperature, the $H_2$ gas is turned off and the furnace is allowed to cool down. (b) Optical microscope image of $Bi_2Se_3$ nanostructures in 14-15.5 cm region. A nanowire (approximately 20 μm long) can be observed in upper right hand corner. (c) SEM image of nanostructures on a glass substrate. (d) SEM image of nanostructures on a silicon substrate. (e) Optical microscope image of nanostructures on mica.**

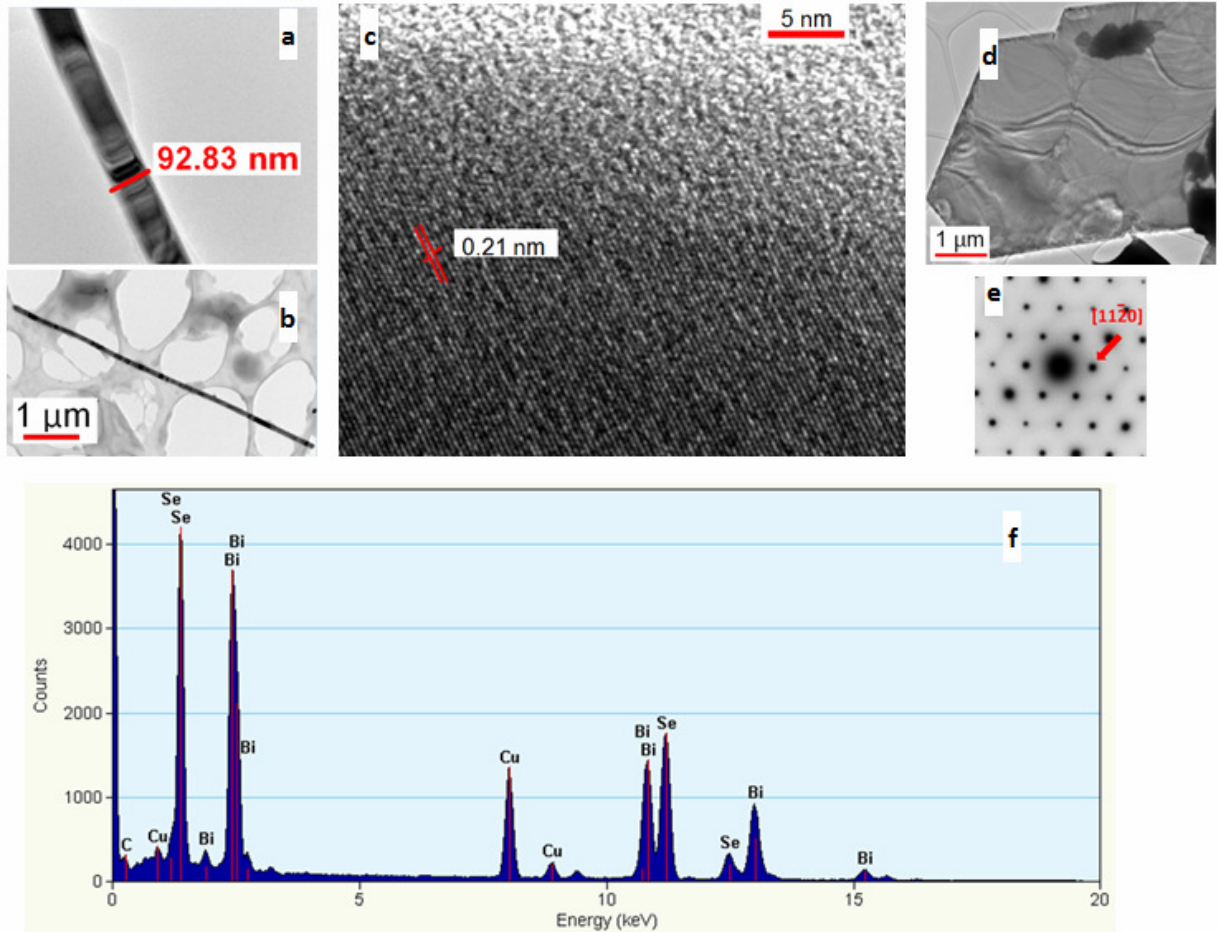

**Figure 2. TEM and structural analysis. (a) A TEM image of a Bi$_2$Se$_3$ nanowire, with a diameter of 93 nm. (b) A TEM image of another nanowire that is 6 μm long and 100 nm in diameter. (c) High-Resolution TEM image of the nanowire shown in (b). Lattice spacing is found to be 0.21 nm in accordance with Bi$_2$Se$_3$ lattice. (d) TEM image of a typical nanoplatelet. (e) SAED of the nanoplatelet shown in (d). Analysis of SAED shows hexagonal structure with the radial pattern corresponding to the [110] direction. (f) EDS of a nanoplatelet shows clear Bi and Se peaks. The C peak is most likely due to contamination during the measurement process and the Cu is from the copper TEM grid on which the structure was measured.**

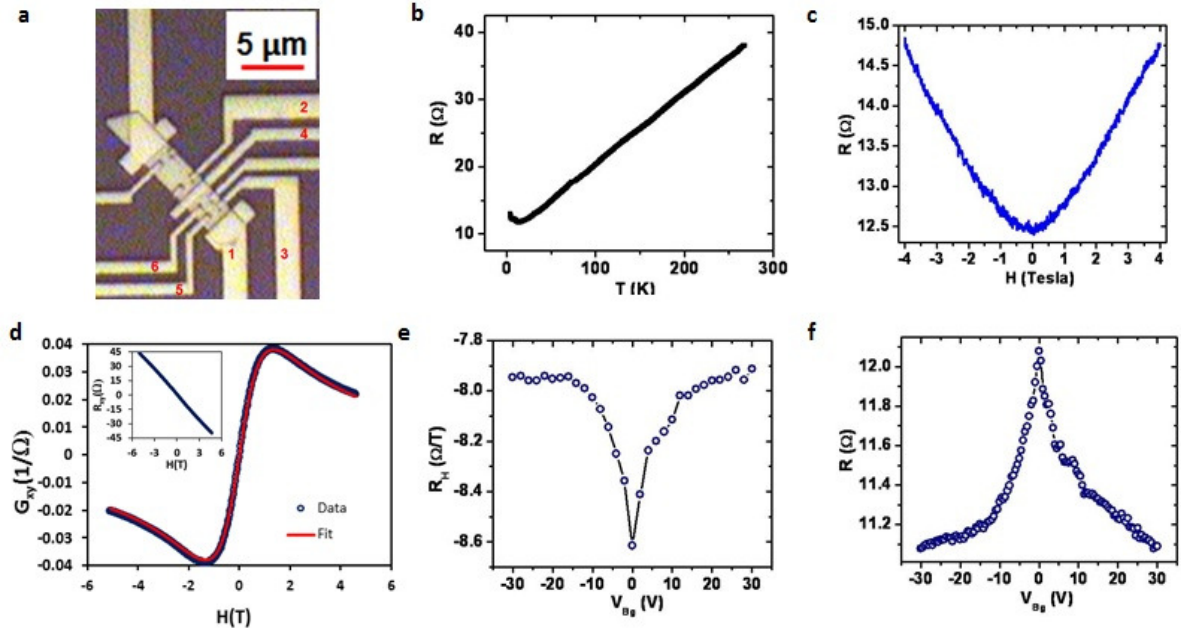

**Figure 3. Bi$_2$Se$_3$ Hall-bar nanoribbon device and transport measurement data. (a) Image of the nanoribbon device with paladium leads. Current is driven through the leads 1 and 2. Longitudinal voltage was measured between leads 3 and 4, and the Hall voltage was measured between leads 3 and 5. The separation between the voltage leads 3 and 4 is 500nm. (b) The longitudinal resistance as a function of temperature. With decreasing temperature the resistance decreases down to 14K. Below 14 K, the resistance shows an upturn as the temperature is decreased further. (c) Longitudinal magnetoresistance measurements show increase of the resistance in increasing magnetic field. (d) Hall conductance data along with a fit to the two-carrier conductance model (see Supporting Information for details on the fit). Hall resistance as a function of magnetic field is shown in the inset. (e) Hall coefficient as a function of back-gate voltage shows dependence on the back gate. (f) The resistance as a function of back-gate voltage also shows that the carrier**

concentration can be controlled by the gate voltage. Measurements shown in (c-f) were done at 4.2 K.

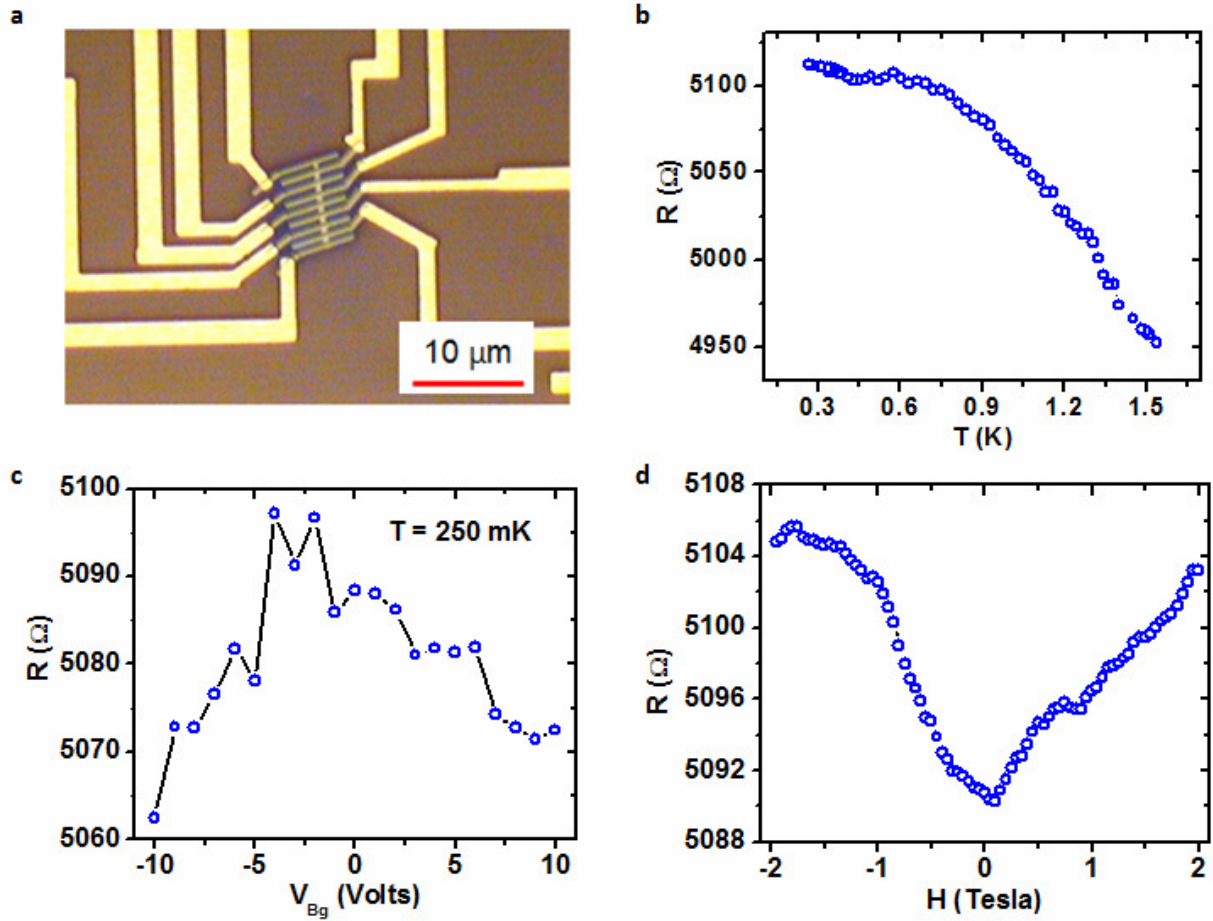

**Figure 4. Nanowire Bi$_2$Se$_3$ device and transport measurements. (a) An image of the device with platinum leads. FIB was used to remove the oxide layer on the nanoribbon and to deposit platinum leads. Platinum leads were connected to bonding pads using palladium. Leads 1 and 2 were used as current leads, and the voltage was measured on leads 3 and 4. (b) R vs. T shows increasing resistance with decreasing temperature. (c) Resistance as a function of back gate voltage shows that the carrier concentration can be controlled by the gate voltage. (d) Magnetoconductance as a function of magnetic field (open circles) shows**

that the conductance decreases as a function of magnetic field. A fit to WAL expression (solid red line) is also shown (see Supplementary Information for more details on the fit).

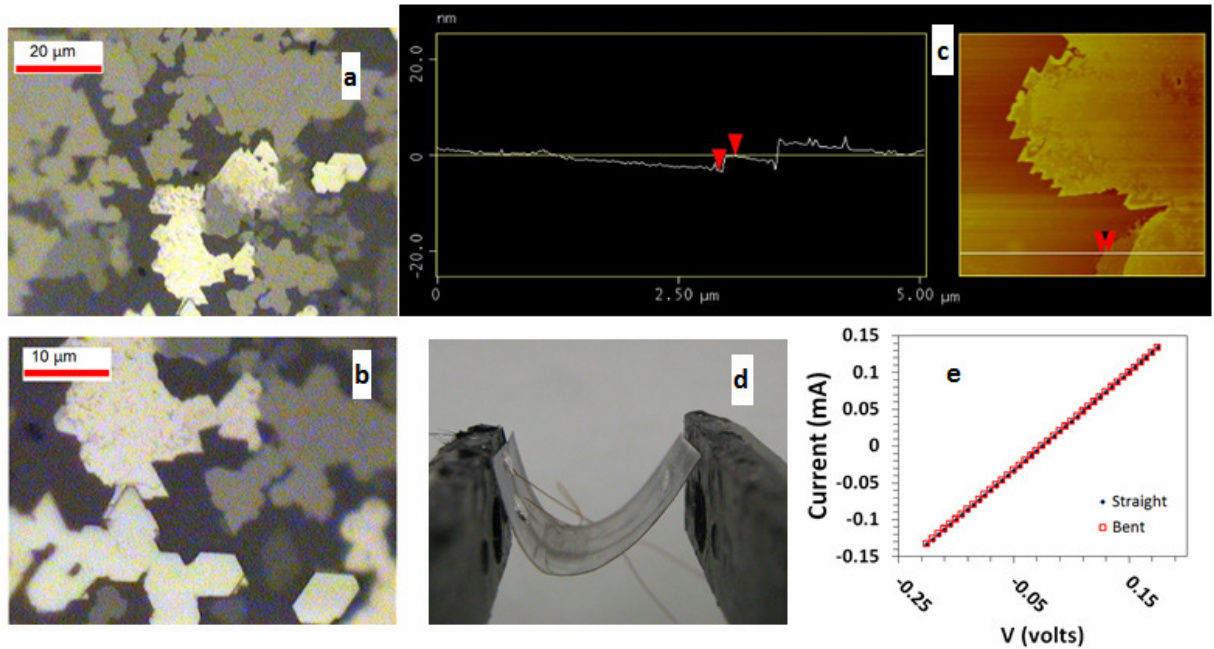

**Figure 5. Optical microscope and atomic force microscope (AFM) images of Bi$_2$Se$_3$ on mica substrates and transport measurements of composite-nanoplatelet films. (a-b) Optical microscope images of the layering of Bi$_2$Se$_3$ nanoplatelets in hexagonal and triangular shapes. (c) AFM image of nanostructures grown on mica in the 14-16 cm temperature region. The height analysis in the thinnest area (between points denoted by red markers) gives a height difference of approximately 3 nm, which is equivalent to the c-axis height of Bi$_2$Se$_3$ unit cell (three quintuple layers) (d) A photograph of a flexible mica substrate with a Bi$_2$Se$_3$ film grown on top and bent using a vice. Electrodes are indium pads, attached by silver paint. (e) I-V characteristics of the Bi$_2$Se$_3$ film on the straight mica substrate (black**

**circles) and on the bent mica substrate (red open squares). Change in the resistance was minimal: the average resistance on straight mica was 1.499 kΩ, and the average resistance on bent mica was 1.501kΩ.**

ACKNOWLEDGMENTS

We sincerely thank Tyrel McQueen for useful suggestions. This work is supported by the National Science Foundation under DMR-1106167 and DGE-1232825 (J. T. M.). N. M. would like to acknowledge the hospitality of the Aspen Center for Physics.

REFERENCES

(1) M. Z. Hasan and C. L. Kane, *Rev. Mod. Phys.* **82**, 3045 (2010).

(2) J. E. Moore, *Nature* **464**, 194 (2010).

(3) X.-L. Qi and S.-C. Zhang, *Rev. Mod. Phys.* **83**, 1057 (2011).

(4) L. Fu and C. L. Kane, *Phys. Rev. B* **76**, 045302 (2007).

(5) H. Zhang, C.-X. Liu, X. Qi, X., Dai, Z. Fang and S.-C. Zhang, *Nat. Phys.* **5**, 438 (2009).

(6) J. E. Moore and L. Balents, *Phys. Rev. B* **75**, 121306(R) (2007).

(7) Oleg V. Yazyev, Joel E. Moore and Steven G. Louie, *Phys. Rev. Lett.* **105**, 266806 (2010).

(8) S. Vishveshwara, *Nat. Phys.* **7**, 450 (2011).

(9) B. Seradjeh, J. E. Moore and M. Franz, *Phys. Rev. Lett.* **103**, 066402 (2009).

(10) L. Fu and C. L. Kane, *Phys. Rev. Lett.* **100**, 096407 (2008).

(11) A. R. Akhmerov, J Nilsson and C. W. J. Beenakker, *Phys. Rev. Lett.* **102**, 216404 (2009).

(12) Y. S. Kim, M. Brahlek, N. Bansal, E. Edrey, G. A. Kapilevich, K. Iida, M. Tanimura, Y. Horibe, S. W. Cheong and S. Oh, *Phys. Rev* **B 84**, 073109 (2011).

(13) G. Zhang, Q. Huajun, J. Teng, J. Guo, Q. Guo, X. Dai, Z. Fang and K Qu, *Appl. Phys. Lett.* **95**, 053114 (2009).

(14) Y. Takagaki, U. Jahn, M. Ramsteiner and K.-J. Friendland, *Semicond. Sci. Technol.* **26**, 085031 (2011).

(15) D. Hsieh, Y. Xia, D. Qian, L. Wray, J. H. Dil, F. Meier, J. Osterwalder, L. Patthey, J. G. Checkelsky, N. P. Ong, A. V. Federov, H. Lin, A. Bansil, D. Grauer, Y. S. Hor, R. J. Cava, and M. Z. Hasan, *Nature* **460**, 1101 (2009).

(16) Y. Xia, D. Qian, D.Hsieh, L. Wray, A. Pal, H. Lin, A. Bansil, D Grauer, Y. S. Hor, R. J. Cava and M. Z. Hasan, *Nat. Phys.* **5**, 398 (2009).

(17) D. Kong, J. J Cha, K. Lai, H. Peng, J. G. Analytis, S. Meister, Y. Chen, H.-J. Zhang, I. R. Fisher, Z.-X Shen, and Y. Cui, *ACS Nano* **5** 4698 (2011).

(18) J. Zhang, Z. Peng, A. Soni, Y. Zhao, Y. Xiong, B. Peng, J. Wang, M. S. Dresselhaus and Q. Xiong, *Nano. Lett.* **11**, 2407 (2011).

(19) X. Qiu, C. Burda, R. Fu, L. Pu, H. Chen and J. Zhu, *J. Am. Chem. Soc.* **126**, 16276 (2004).

(20) J. G. Checkelsky, Y. S. Hor, R. J. Cava and N. P. Ong, *Phys. Rev. Lett.* **106**, 196801 (2011).

(21) S. S. Hong, W. Kundhikanjana, J. J. Cha, K. Lai, D. Kong, S. Meister, M. A. Kelly, Z.-X. Shen and Y. Cui, *Nano. Lett.* **10**, 3118 (2010).

(22) D. Kong, J. C. Randel, H. Peng, J. J. Cha, S. Meister, K. Lai, Y. L. Chen, Z. X. Shen, H. C. Manoharan and Y. Cui, *Nano Lett*. **10**, 329 (2010).

(23) D. Kong, W. Dang, J. J. Cha, H. Li, S. Meister, H. Peng, Z. Liu and Y. Cui, *Nano Lett.* **10** 2245 (2010).

(24) L. D. Algeria, M. D. Schroer, A. Chatterjee, G. R. Poirer, M. Pretko, S. K. Patel and J. R. Petta, *Nano Lett*. **12**, 4711 (2012).

(25) Y. Zhang, K. He, C.-Z. Chang, C.-L. Song, L.-L. Wang, X. Chen, J.-F. Jia, Z. Fang, X. Dai, W.-Y. Shan, S.-Q. Shen, Q. Niu, X.-L. Qi, S.-C. Zhang, X.-C. Ma and Q.-K. Xue, *Nat. Phys.* **6**, 584 (2010).

(26) S. Cho, N. P. Butch, J. Paglione and M. S. Fuhrer, *Nano.Lett.* **11**, 1925 (2011).

(27) H. Steinberg, D. R. Gardner, Y. S. Lee, and P. Jarillo-Herrero, *Nano. Lett.* **10**, 5032 (2010).

(28) H. Peng, W. Dang, J. Cao, D. Wu, W. Zheng, H. Li, Z. X. Shen and Z. Liu, *Nat. Chem.* **4**, 281 (2012).

(29) H. Tang, D. Liang, L. J. Richard, and X. P. A. Gai, *ACS Nano* **5**, 7510 (2011).

(30) H. Peng, K. Lai, D. Kong, S. Meister, Y. Chen, X.-L. Qi, S.-C. Zhang, Z.-X. Shen and Y. Cui, *Nat. Mat.* **9**, 225 (2010).

(31) H. Steinberg, J.-B. Laloë, V. Fatemi and P. Jarillo-Herrero, *Phys. Rev. B* **84**, 233101 (2011).

(32) S. S. Hong, J. J. Cha, D. Kong and Y. Cui, *Nature Communications* , DOI: 10.1038/ncomms1771 (2012).

(33) X. He, T. Guan, X. Wang, B. Feng, P. Cheng, L. Chen, Y. Li and K. Wu, *Appl. Phys. Lett.* **101**, 123111 (2012).

(34) N. Bansal, Y. S. Kim, M. Brahlek, E. Edrey and S. Oh, *Phys. Rev. Lett.* **109**, 116804 (2012).

(35) J. Chen, H. J. Qin, F. Yang, J. Liu, T. Guan, F. M. Qu, G. H. Zhang, J. R. Shi, X. C. Xie, C. L. Yang, K. H. Wu, Y. Q. Li and L. Lu, *Phys. Rev. Lett.* **105**, 176602 (2010).

(36) H. Lind, S. Lidin, and U. Häussermann, *Phys. Rev*. **B 72**, 184101 (2005).

(37) I. Garate and L. I. Glazman, *Phys. Rev*. **B 86**, 035422 (2012).

(38) J. G. Checkelsky, Y. S. Hor, M.-H. Liu, D.-X. Qu, R. J. Cava, and N. P. Ong, *Phys. Rev. Lett.* **103**, 246601 (2009).

Supporting Information for

# Catalyst-Free Synthesis of High Purity Topological Insulator $Bi_2Se_3$ Nanostructures

*Jerome T. Mlack, Atikur Rahman, Gary L. Johns, Kenneth J.T. Livi, Nina Markovic*

## A. Images of structures grown without $H_2$ gas

Optical images of nanostructures grown without $H_2$ as an assisting carrier gas are shown in figure S1. The growth is performed using the same method as outlined in the main text, but without $H_2$ as a secondary carrier gas. There are far fewer nanostructures and they are significantly smaller than those obtained by the catalyst free method with $H_2$ gas.

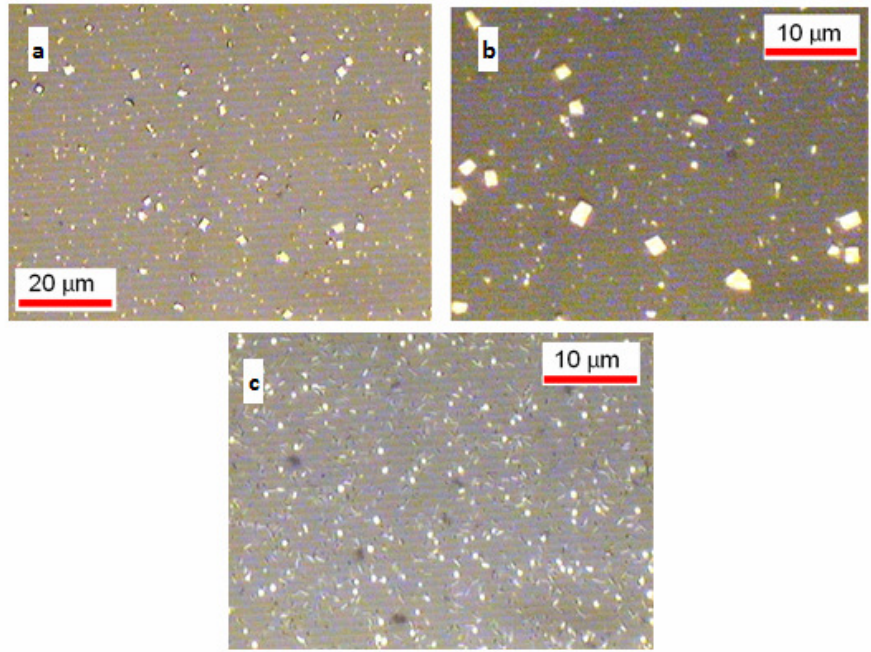

**Figure S1. (a-c)** Optical images of nanostructures grown on glass slides using the VS catalyst-free method *without* $H_2$ as a carrier gas. There are far fewer structures and the structures are smaller in size compared to the growth results in the presence of $H_2$ gas.

### B. Images of growth results at varying distances from the center of the furnace

At different distances from the center of the furnace, the temperature gradient determines the type of the resulting structures. Figure S2 shows the growth results on glass slides at varying distances from the center, with a growth temperature set at 590° C, the argon flow rate set to 120 sccm and the hydrogen flow rate set to 3 sccm. Figure S2a shows results on a glass slide centered at 11.5 cm with no noticeable structure growth. Figure S2b was centered at 14 cm and shows growth of nanowires and flakes. At 16.5 cm there are distinct nanowires and ribbons growing upward from the substrate as showing figure S2c. Figures S2d and S2e are both from a slide centered at 19.3 cm from the center and are just outside the furnace in the quartz tube. Although there are large crystals in figure S2d, they melt at 300°C, and are most likely selenium crystals.

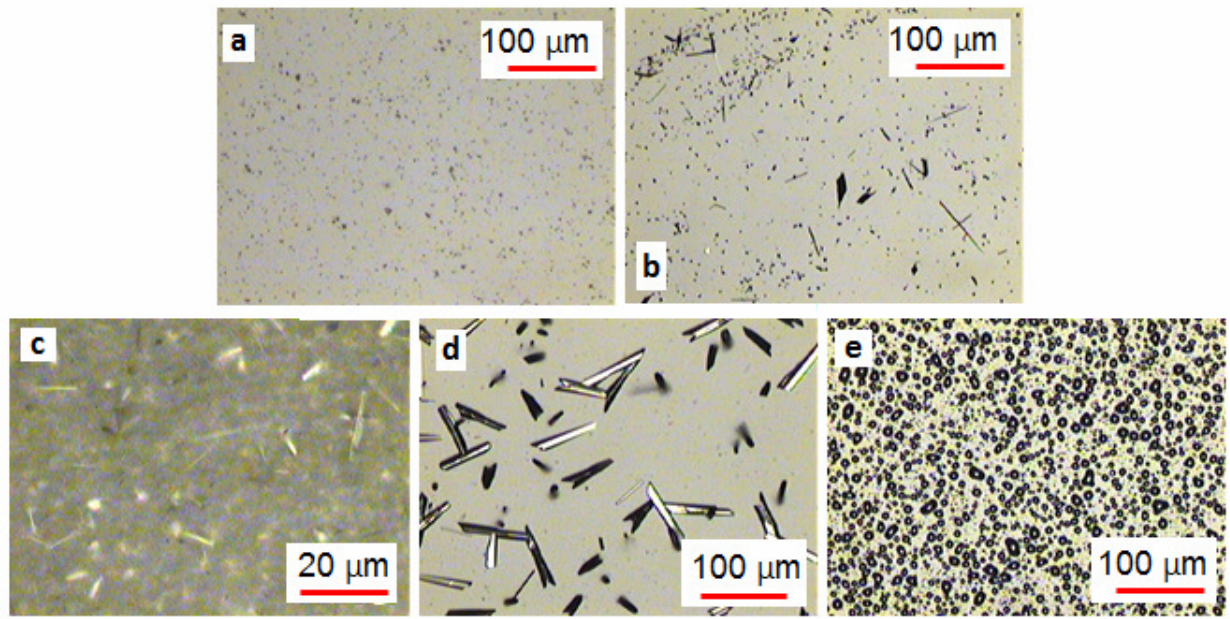

**Figure S2. Optical microscope images of growth results at varying distance from the center of the furnace. He following growth parameters were used: furnace temperature of 590 °C, argon flow of 120 sccm, and hydrogen flow of 3 sccm. (a) Growth results in the region centered at 11.5 cm from the center shows no structures. (b) At 14 cm from the center, we observe wires and flakes. (c) At 16.5cm from the center, clear wires and ribbons grow up from the substrate surface. (d-e) These images are from the region centered at 19.3 cm from the center, which is just outside the furnace. In (d) although there are large crystals, they will melt at 300° C.**

## C. Images of Nanostructure growth on Silicon Oxide

Silicon wafers with 300nm of silicon oxide were cut to 7.5 mm x 7.5 mm, cleaned in the same manner as the glass slides before growth, and placed on top of the glass slides during growth. The results are shown in figure S3. All images show the nanostructures as they are grown on the substrate. Wires and ribbons were found to be abundant in the 14-15 cm region similar to the results of the growth on glass slides. Figure S3 (a) is an optical image of the nanostructures. Nanowires similar to the one shown in the SEM image figure S3 (b) were found across the sample. Some ribbons and flakes were thin enough such that they were translucent under the SEM, such as the nanoribbons shown in figures S3(c) and S3(d).

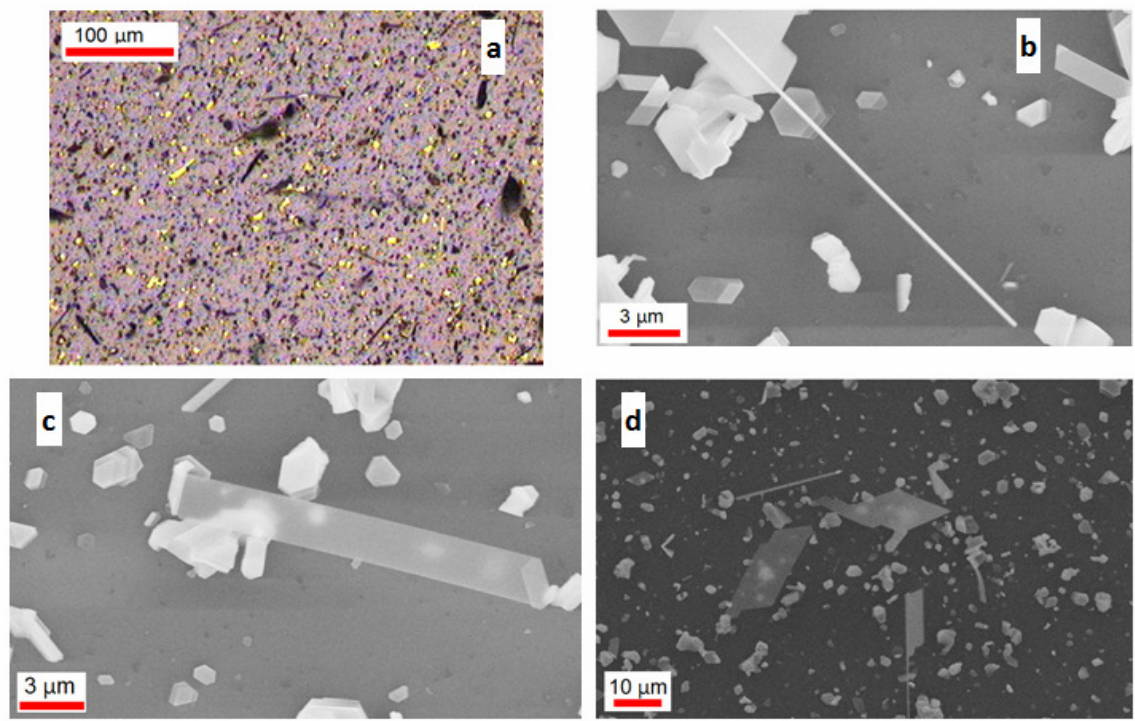

**Figure S3. (a) Optical image of typical $Bi_2Se_3$ nanostructures grown on silicon substrate. Growth was performed in the same manner as that for glass slide substrates. (b) SEM image of $Bi_2Se_3$ nanowire of diameter 240 nm. (c-d) typical SEM images of various as-grown nanostructures. Some nanoribbons were thin enough that they appear translucent under the SEM electron beam.**

## D. Two-Carrier Model

If the sample has both bulk and surface conduction then Rxy will be non-linear. This non-linearity is due to two or more conduction channels that have similar but distinct carrier densities and mobilities. In figure 3d we fit Gxy using a two-carrier model[34]. The model has two free parameters and is given as:

$$G_{xy}(B) = eB\left(\frac{c_1\mu_1 - c_2}{(\mu_1/\mu_2 - 1)(1+\mu_2^2 B^2)} + \frac{c_1\mu_2 - c_2}{(\mu_2/\mu_1 - 1)(1+\mu_1^2 B^2)}\right)$$

Where $\mu_1$ and $\mu_2$ are the mobilities of the two carriers and are the free parameters in the fitting.

The constants and $c_2$ are determined directly from the data. $c_1$ is the measured conductance at zero field and $c_2$ linear slope of the hall conductance near zero field:

$$c_1 = \frac{G_{xx}(0)}{e} \qquad c_2 = \lim_{B \to 0} \frac{G_{xy}(B)}{eB}$$

The two carrier densities are calculated by:

$$n_1 = \frac{c_1\mu_2 - c_2}{\mu_1\mu_2 - \mu_1^2} \qquad n_2 = \frac{c_1\mu_1 - c_2}{\mu_1\mu_2 - \mu_2^2}$$

In order to fit the conductance data, we convert the measured resistance into conductance using the conductance tensor:

$$G_{xy}(B) = -\frac{R_{xy}^2}{R_{xy}^2 + R_{xx}^2}$$

The fitting gives carrier concentrations of $n_1 = 6.0e+13$ $1/cm^2$ and $n_2 = 7.8e+12$ $1/cm^2$. The concentration values are in agreement with those observed by N. Bansal et. al[34].

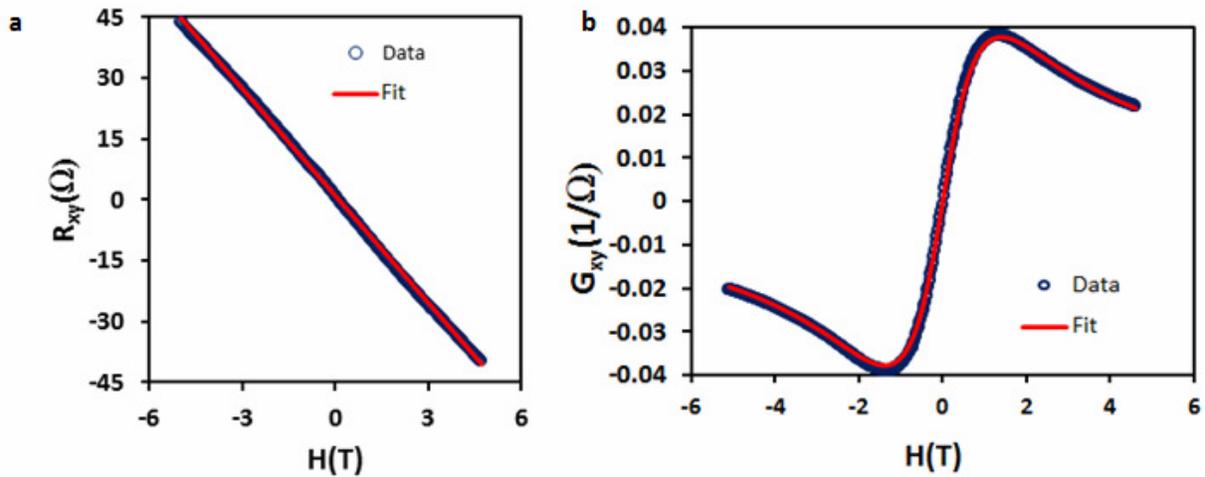

**Figure S4. Fitting the two-carrier model to the data obtained on device shown in Figure 3(a) in the main text. (a) Rxy data with linear fit. Careful analysis of the data shows that it is not quite linear. (b) Transverse magneto conductance was fit using the two-carrier model[34]. The fit yields carrier concentrations of $n_1 = 6.0e+13$ $1/cm^2$ and $n_2 = 7.8e+12$ $1/cm^2$.**

### E. Weak antilocalization (WAL)

One of the hallmarks of a topological insulator is strong spin-orbit coupling which leads to WAL. WAL arises due to the destructive interference of electron scattering paths, causing an overall increase of the average conductance of the sample in the absence of a magnetic field. The magnetic field breaks the time-reversal symmetry and removes the WAL correction, resulting in negative magnetoconductance. The quantum correction to the conductance arising from WAL is given by [1]

$$\Delta G(B) = \alpha \frac{e^2}{\pi \hbar}\left[\ln\frac{B_0}{B} - \psi\left(\frac{1}{2}+\frac{B_0}{B}\right)\right]$$

where $B_0 = \hbar/4el_0^2$ is the dephasing magnetic field determined by the phase coherence length $l_0$, $\psi$ is the digamma function, and $\alpha = -1/2$ if the system exhibits strong spin-orbit interactions.

Fitting the magnetoconductance to equation 1 with $\alpha=-0.5$ yields a phase coherence length of 15nm. The contribution of bulk carriers to the conduction was not taken into account for this fit.

(39) 1. Hikami, S.; Larkin, A. I.; Nakaoka, Y. *Prog. Theor. Phys.*, **63**, 707-710 (1980).